\documentclass[notitlepage,nofootinbib]{revtex4}
\usepackage[utf8]{inputenc}
\usepackage{bbold}

\usepackage{natbib}
\usepackage{graphicx}
\usepackage{epsfig}
\usepackage{amsmath,amssymb}
\input{epsf}
\usepackage{psfrag}

\usepackage[usenames,dvipsnames]{color}

\newcommand{\beq}{\begin{eqnarray}}
\newcommand{\eeq}{\end{eqnarray}}
\newcommand{\Slash}[1]{{\ooalign{\hfil/\hfil\crcr$#1$}}}

\newcommand{\nn}{\nonumber \\}

\usepackage{amsmath,color}
\usepackage{amssymb}
\usepackage{bm}% bold math
\usepackage{graphicx}
\usepackage{multirow}

\begin{document}

\title{Nucleon electric dipole moment from polarized deep inelastic scattering}

\author{Yoshitaka Hatta}

\affiliation{Brookhaven National Laboratory, Upton, NY 11973, USA}

\affiliation{RIKEN BNL Research Center, Brookhaven National Laboratory, Upton, NY 11973, USA}

\begin{abstract}
In a previous paper \cite{Hatta:2020ltd}, we have pointed out the connection between the $CP$-odd three-gluon (Weinberg) operator and certain twist-four corrections in polarized deep inelastic scattering. Based on this observation, we give a numerical estimate of the electric dipole moment of the proton and neutron induced by the Weinberg operator. Our result tends to be smaller than the previous estimates based on QCD sum rules by a factor of about 3 or more, though they are not entirely inconsistent considering theoretical uncertainties.

\end{abstract}

\maketitle

\section{Introduction}

Permanent electric dipole moments (EDM) of elementary particles, the nucleons, nuclei and atoms are an unambiguous signal of $CP$-violation in Nature that has profound implications for the evolution and composition of the  universe. While being definite predictions of the Standard Model (SM) and  models beyond the SM (BSM),  EDMs in fundamental particles have never been observed so far despite  decades of  searches by many experimental collaborations (see reviews  \cite{Chupp:2017rkp,Yamanaka:2017mef} and references therein).  Currently, only  upper bounds for various systems have been determined, and the  race is on worldwide to lower these bounds by several orders of magnitude in the next decades to come. The  sensitivity of measurement in the present experiments has already reached the region predicted by many BSM scenarios, thereby providing stringent constraints on model parameters.

Typically, some  BSM physics at the TeV scale or above induces, at lower energies,  high-dimensional $CP$-violating operators into the QCD Lagrangian via loop effects. From that point on, it becomes entirely a QCD problem  to determine the EDM of hadrons and nuclei generated by these operators. Of particular interest is the purely gluonic, dimension-six operator called the  Weinberg operator \cite{Weinberg:1989dx} 
\beq
{\cal O}_W = gf^{abc}\tilde{F}_{\mu\nu}^a F^{\mu\sigma}_b F^\nu_{c \sigma}.
\eeq
Unlike the other  $CP$-violating operators which involve the quark fields, ${\cal O}_W$ does not receive  suppression due to the current quark masses. Moreover, unlike the topological operator $\tilde{F}^{\mu\nu}F_{\mu\nu}$,  ${\cal O}_W$ is free of the strong $CP$ problem and does not affect the vacuum structure of QCD. These considerations make ${\cal O}_W$ a relatively `clean' source of hadronic EDMs.   However, purely-gluonic high-dimensional operators such as ${\cal O}_W$ are quite difficult to deal with theoretically. The most promising approach, the lattice QCD simulation, is not yet at the practical level despite much  progress in recent years \cite{Cirigliano:2020msr,Rizik:2020naq}. More phenomenological approaches \cite{Bigi:1990kz,Demir:2002gg,Haisch:2019bml,Yamanaka:2020kjo} are subject to large theoretical uncertainties, but  at least they can provide concrete numbers that are reasonable in terms of magnitude and can be used as a reference. 

In this paper, we aim to improve on the phenomenological approach  by suggesting a new method to evaluate an important part of the nucleon EDM, namely, the so-called `one-nucleon reducible' diagrams in the classification of \cite{Bigi:1990kz}. Our approach is based on the observation \cite{Hatta:2020ltd} that ${\cal O}_W$ mixes with a certain twist-four, quark-gluon operator known in the context of higher-twist corrections in polarized deep inelastic scattering (DIS). We show that the matrix element of the latter operator directly affects the nucleon EDM through the one-nucleon reducible diagrams. Moreover, it can be extracted from the data of the existing and future polarized DIS experiments such as the Electron-Ion Collider (EIC) \cite{Proceedings:2020eah} in the US. This opens up a novel, unexpected connection between traditional QCD spin physics and the physics of the nucleon EDM. As a demonstration of our method, we employ a phenomenological value of the matrix element and make a simple estimate of the proton and neutron EDMs.

\section{Nucleon EDM from the Weinberg operator}

Throughout this paper, we assume that the Weinberg operator is induced in QCD as a low-energy effective action  due to some beyond the standard model (BSM) physics \cite{Weinberg:1989dx}
\beq
\Delta {\cal L}=w\int d^4x\,  gf^{abc}\tilde{F}_{\mu\nu}^a F^{\mu\sigma}_b F^\nu_{c \sigma} \equiv w\int d^4x {\cal O}_W(x), \label{eff}
\eeq
 where $\tilde{F}_{\mu\nu}=\frac{1}{2}\epsilon_{\mu\nu\alpha\beta}F^{\alpha\beta}$ and $\epsilon^{0123}=+1$.  The coefficient $w$ has dimension $-2$. Different normalizations of the operator are used in the literature. Here we include one factor of the QCD coupling $g$ in ${\cal O}_W$, and our convention is such that $g$ enters the covariant derivative with a positive sign $D^\mu=\partial^\mu+igA^\mu$. This choice is convenient when establishing connections to higher twist operators in the QCD literature. It is also a preferred definition since the sign of the coupling $g$ is indefinite in QCD due to the symmetry $g,A_\mu,F_{\mu\nu} \to -g,-A_\mu,-F_{\mu\nu}$ of the Lagrangian. With our normalization,  $w$ must be even under  $g\to -g$. 
 
\begin{figure}[htbp]
\centering
\includegraphics[width=0.9\textwidth]{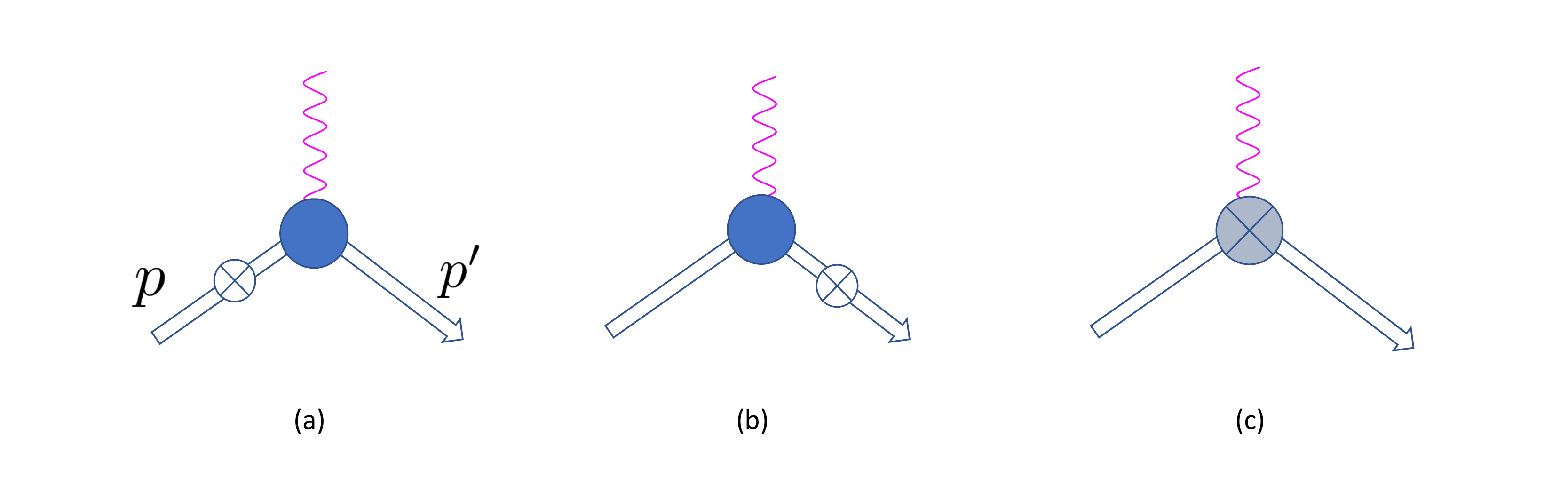}
\caption{Three types of diagrams that contribute to the nucleon EDM. The symbol $\otimes$ denotes the insertion of the Weinberg operator.}
\label{fig1}
\end{figure}

Consider the electromagnetic form factor
 of the nucleon in the presence of $\Delta {\cal L}$
\beq
\langle P'S'|J^\mu_{em}(0)|PS\rangle = \bar{u}(P'S') \left[\gamma^\mu F_1(\Delta^2) + \frac{i\sigma^{\mu\nu}\Delta_\nu}{2m}(F_2(\Delta^2) -i\gamma_5 F_3(\Delta^2)) \right]u(PS) \label{form}
\eeq
where $\Delta^\mu=P'^\mu-P^\mu$ and $m$ is the nucleon mass. $J_{em}^\mu=\sum_f e_f|e| \bar{\psi}_f \gamma^\mu \psi_f$ is the electromagnetic current ($e_f=2/3$ for the $u$-quark, etc). Our conventions are $\gamma_5=-i\gamma^0\gamma^1\gamma^2\gamma^3$ and $\sigma^{\mu\nu}=\frac{i}{2}[\gamma^\mu , \gamma^\nu]$. 
The spin 4-vector $S^\mu$ is defined by 
\beq
2S^\mu = \bar{u}(PS)\gamma_5\gamma^\mu u(PS)
\eeq
with the properties $S\cdot P=0$ and $S^2=-m^2$. 
 The $CP$-violating $F_3$ form factor is nonzero due to  the effective action (\ref{eff}). Its value at vanishing momentum transfer is related to the EDM of the proton ($p$) and neutron ($n$)
 \beq
 \frac{F^{p,n}_3(0)}{2m}=d_{p,n}.
 \eeq
Finding the relation between $w$ and $F_3$ is a complicated nonperturbative problem of QCD. The usual argument, put forward in \cite{Bigi:1990kz}, is that diagrammatically there are three types of contributions as depicted in Fig.~\ref{fig1}. The first two diagrams (a,b) are `one-nucleon reducible' and feature the nucleon intermediate state. The last diagram (c) represents the sum of  `one-nucleon irreducible' contributions that are sensitive to physics at shorter distances. It has been argued in \cite{Bigi:1990kz} that the reducible diagrams dominate, or at least provide the correct order of magnitude estimate, see also  \cite{Demir:2002gg,Haisch:2019bml}.   The irreducible contribution has been recently calculated in the quark model in  \cite{Yamanaka:2020kjo}, and the result is  indeed  smaller than the contribution from the reducible  diagrams calculated in  \cite{Demir:2002gg,Haisch:2019bml}.   
 
In the remainder of this section, we shall explicitly evaluate the reducible diagrams and confirm the result \cite{Bigi:1990kz} 
\beq
d_n \approx \mu_n \frac{\langle N|w{\cal O}_W|N\rangle}{\bar{u}_N i\gamma_5 u_N}, \label{amb}
\eeq
where $\mu_n$ is the  magnetic moment of the neutron.  While such an exercise may seem mundane (the derivation was omitted in  \cite{Bigi:1990kz}), it still has the merit of clarifying  subtleties  that are not always  articulated in the literature.
%The exact meaning of (\ref{amb}) is somewhat unclear, partly due to the  obscure notation $\langle N|...|N\rangle$ ($N$ for `nucleon') which does not specify the momentum and spin states.
In particular, both the numerator and denominator of (\ref{amb}) vanish in the forward limit. This may seem   problematic because ${\cal O}_W$ is inserted at zero momentum transfer $\int d^4x {\cal O}_W(x)$, so naively only the forward matrix elements are relevant. In  \cite{Bigi:1990kz}, this issue was  obscured by the notation $\langle N|...|N\rangle$ ($N$ for `nucleon') which does not specify the momentum and spin states.\footnote{And also by a typo in Eq.~(10) of Ref.~\cite{Bigi:1990kz}: $\tilde{O}_{(6)}(0)J_\mu^{em}(x)$ should read $\tilde{O}_{(6)}(x)J_\mu^{em}(0)$.} We shall instead use the more transparent notation $\langle P'S'|...|PS\rangle$ throughout.  We would like to also clarify  whether $\mu$ in (\ref{amb}) is the total or anomalous magnetic moment, a question relevant when applying (\ref{amb}) to the proton case. 

 Since $w$ is small, it is enough to keep only the linear terms in $w$. The $CP$-violating part of the form factor then reads   
\beq
\langle P'S'|J^\mu_{em}(0)|PS\rangle &\sim&iw\int d^4x \langle P'S'|{\rm T}\{J^\mu_{em}(0){\cal O}_W(x)\}|PS\rangle \nn
&=& iw\int d^4x \langle P'S'|\theta(x^0) {\cal O}_W(x)J_{em}^\mu(0)+\theta(-x^0)J^\mu_{em}(0){\cal O}_W(x)|PS\rangle. \label{last}
\eeq
One can isolate the reducible diagrams  by inserting the complete set 
\beq
1=\sum_{S''}\int \frac{d^3k}{(2\pi)^32E_{k}}|kS''\rangle\langle kS''| + \cdots
\eeq
with $k^0=E_k=\sqrt{m^2+\vec{k}^2}$ 
and keeping only the single nucleon state. The contributions from resonances $N^*$ and multi-particle states $\pi N, \pi\pi N,\cdots$ should be regarded as  part of the irreducible diagram  Fig.~\ref{fig1}(c). This gives 
\beq
&&\int d^4x \langle P'|\theta(x^0) {\cal O}_W(x)J_{em}^\mu(0) |P\rangle \nn
&&\approx  \sum_{S''}\int_0^\infty dx^0 \int d^3x \int \frac{d^3k}{(2\pi)^32E_k} e^{ix(P'-k)}\langle P'|{\cal O}_W(0)|kS''\rangle\langle kS''|J_{em}^\mu(0)|P\rangle
\nn
%&&\approx \sum_{S''} \int_0^\infty dx^0 \int d^3x\int \frac{d^3k}{(2\pi)^32E_k}  e^{ix(P'-k)} \langle P'|{\cal O}_W(0)|k\rangle\langle k|J_{em}^\mu(0)|P\rangle  \nn 
&& \left.=\sum_{S''} \frac{i}{2E_k(E_{P'}-E_k+i\epsilon)} \langle P'S'|{\cal O}_W(0)|kS''\rangle\langle kS''|J_{em}^\mu(0)|PS\rangle  \right|_{\vec{k}=\vec{P}'} \label{ls}
\eeq
 The electromagnetic form factor in (\ref{ls}) should be that of $CP$-conserving theory. 
The matrix element of the  Weinberg operator can be parameterized as  \cite{Hatta:2020ltd}
\beq
\langle P'S'|{\cal O}_W(0)|kS''\rangle  &=&4m^3E((P'-k)^2))\bar{u}(P'S')i\gamma_5 u(kS'') \nn  &=&2m^2E((P'-k)^2) (P'-k)_\nu \bar{u}(P'S')\gamma^\nu i\gamma_5 u(kS'') \label{wein}
\eeq
In (\ref{ls}), there is a pole at $E_{P'}=E_k$ and in (\ref{wein}) there is a zero at $P'^\nu=k^\nu$. Thus, the limit $k\to P'$ has to be taken carefully.  
Adding the contribution from the second term in (\ref{last}) and using 
\beq
\sum_{S''} u(kS'')\bar{u}(kS'')&=&\Slash k +m\nn 
\bar{u}(P'S')\gamma_5 (\Slash k +m) &=& -(P'_\nu-k_\nu)\bar{u}(P'S')\gamma_5\gamma^\nu \nn 
(\Slash k +m)\gamma_5 u(PS) &=& (k_\nu-P_\nu)\gamma^\nu \gamma_5 u(PS),
\eeq
 one finds the relation 
\beq
&& 2iwm^3 E(0) \Biggl( \lim_{k\to P'}\frac{P'_\nu-k_\nu}{E_{P'}(E_{P'}-E_k)} \bar{u}(P'S')\gamma_5\gamma^\nu  \left[\gamma^\mu F_1 + \frac{i\sigma^{\mu\nu}\Delta_\nu}{2m}F_2 \right] u(PS) \nn 
&&  \qquad \qquad  + \lim_{k\to P} \frac{k_\nu-P_\nu}{E_{P}(E_{k}-E_P)} \bar{u}(P'S')\left[\gamma^\mu F_1 + \frac{i\sigma^{\mu\nu}\Delta_\nu}{2m}F_2 \right]\gamma^\nu\gamma_5  u(PS) \Biggr)\nn 
&&=\bar{u}(P'S')  \frac{\sigma^{\mu\nu}\Delta_\nu}{2m} \gamma_5F_3(\Delta^2) u(PS). \label{rhs}
\eeq
Since the EDM is 
 proportional to $F_3$ at $\Delta=0$, one can write $\bar{P}=\frac{P'+P}{2}$  and keep only linear terms in $\Delta=P'-P$. Further, one can set $S=S'$, and take the nonrelativistic limit in which $\vec{P},\vec{P}',\vec{\Delta}$ are all small. The right hand side of (\ref{rhs}) then reduces to  
\beq
\bar{u}(\bar{P}S)  \sigma^{\mu\nu}\gamma_5\Delta_\nu u(\bar{P}S) = \frac{-2i}{m}(\bar{P}^\mu S^\nu - \bar{P}^\nu S^\mu)\Delta_\nu 
%&=& \frac{-2i}{m}S\cdot \Delta \bar{P}^\mu  \nn
%&\approx& \frac{2i}{m}\vec{S}\cdot\vec{\Delta}\bar{P}^\mu
\approx 2i\vec{S}\cdot \vec{\Delta}g^{\mu 0}
\eeq
where we used $\bar{P}\cdot \Delta=0$ and the fact that  $S^0=\frac{\vec{S}\cdot \vec{P}}{E_P}$ %(which follows from $S\cdot P=0$) 
is small in the nonrelativistic regime. 
%That is, we neglect  quadratic terms in spatial momentum $\vec{P},\vec{P}',\vec{\Delta}$. 
 
% The spinor part can be approximated as 
%\beq
%&&\bar{u}(P'S') \left[\gamma^\mu F_1(\Delta^2) + \frac{i\sigma^{\mu\nu}\Delta_\nu}{2m}F_2(\Delta^2)  \right]u(PS) \nn 
%&&= \frac{\bar{P}^\mu}{m}F_1\bar{u}(P'S')u(PS) +\bar{u}(P'S') \left[\frac{i\sigma^{\mu\nu}\Delta_\nu}{2m}(F_1+F_2)  \right]u(PS) \nn
%&& \approx 2\bar{P}^\mu F_1 -\frac{i(F_1+F_2)}{m^2}\epsilon^{\mu\nu\rho\lambda}\Delta_\nu \bar{P}_\rho S_\lambda \label{first}
%\eeq
To evaluate the left hand side of (\ref{rhs}),  some kind of regularization is necessary. We temporarily treat the intermediate state as a `resonance' and modify its mass as $m\to \tilde{m}=m+\epsilon$. It then follows that $\vec{P}'=\vec{k}$ but $E_{P'} \neq \tilde{E}_k$, so that only the $\nu=0$ component remains. After cancelling $E_{P'}-\tilde{E}_k$ in the numerator and denominator, we take the limit $\epsilon\to 0$. In this way,  the brackets in (\ref{rhs})  reduces to 
\beq
&& \bar{u}(P'S)\Biggl( \frac{1}{E_{P'}}\gamma_5\gamma^0 \left[\gamma^\mu F_1 + \frac{i\sigma^{\mu\nu}\Delta_\nu}{2m}F_2 \right]  +  \frac{1}{E_{P}}  \left[\gamma^\mu F_1 + \frac{i\sigma^{\mu\nu}\Delta_\nu}{2m}F_2 \right]\gamma^0\gamma_5  \Biggr)u(PS) \nn
&& \approx \frac{1}{m}\bar{u}(P'S')\left( 2g^{\mu 0}F_1 \gamma_5 + \frac{iF_2}{2m} \gamma_5 2i(g^{0\mu}\gamma^\nu -g^{0\nu}\gamma^\mu)\Delta_\nu \right)u(PS) \nn 
%&& \approx -\frac{2}{m^2}(F_1+F_2)S\cdot \Delta g^{\mu 0} 
&&\approx \frac{2}{m^2}(F_1+F_2)\vec{S}\cdot \vec{\Delta} g^{\mu 0}
\eeq
%\beq
% &&  \left( \frac{1}{E_{P'}} \bar{u}(P'S')\gamma_5\gamma^0 u(P'S') -\frac{1}{E_{P}} \bar{u}(PS)\gamma_5\gamma^0 u(PS)\right)
%  \nn 
% && =2\left( \frac{S'^0}{E_{P'}} -\frac{S^0}{E_P}\right) 
%  = 2 \left(\frac{\vec{S}'\cdot \vec{P}'}{(E_{P'})^2} -\frac{\vec{S}\cdot \vec{P}}{(E_P)^2}\right) \approx \frac{2}{m^2}\vec{S}\cdot \vec{\Delta}
% \eeq
%again in the nonrelativistic limit. Since this is linear in $\Delta$, I only keep the first term in  (\ref{first}) to get 
We thus arrive at
 \beq
d= \frac{F_{3}(0)}{2m} = 4wE(0)m^2\frac{F_1(0)+F_2(0)}{2m}= 4wE(0) m^2 \mu \label{bigi}
 \eeq
  In the present normalization, $F_1(0)=|e|$ for proton and $F_1(0)=0$ for neutron. 
 Eq.(\ref{bigi}) essentially agrees with (\ref{amb}), but now all the subtleties have been exposed. The present derivation also makes it clear that in the proton case  
 the EDM is proportional to the total magnetic moment $F_1+F_2$, not the anomalous magnetic moment $F_2$.

\section{Operator mixing}

Now that we have the relation (\ref{bigi}), the remaining task is to estimate the QCD matrix element $E(\Delta^2=0)$. The central observation of \cite{Hatta:2020ltd} is that $E$ is related to the matrix element of the following twist-four operator 
\beq
\langle PS|\bar{\psi}g\tilde{F}^{\mu\nu}\gamma_\nu \psi |PS\rangle = -2f_0 m^2 S^\mu \label{same}
\eeq
(We have changed the normalization of $f_0$ by a factor of 2 with respect to \cite{Hatta:2020ltd}.)
More precisely, $E$ and $f_0$ mix under  renormalization group (RG) as follows\footnote{From now on, the argument of $E$ is the RG scale, not the momentum transfer $\Delta^2$. The latter has been set equal to zero.}
\beq
E(\mu_0)-\frac{9N_c^2}{2(3N_c^2+4)} f_0(\mu_0) =\left(\frac{\alpha_s(\mu_0)}{\alpha_s(\mu)}\right)^{\frac{\gamma_W}{\beta_0}} \left(E(\mu)-\frac{9N_c^2}{2(3N_c^2+4)} f_0(\mu) \right)
\eeq
 \beq
 f_0(\mu_0)= \left(\frac{\alpha_s(\mu_0)}{\alpha_s(\mu)}\right)^{\frac{\gamma_4}{\beta_0}}  f_0(\mu), \qquad \gamma_4=  \frac{8C_F}{3}+\frac{2n_f}{3} \label{g4}
\eeq 
\beq
w(\mu_0) = \left(\frac{\alpha_s(\mu_0)}{\alpha_s(\mu)}\right)^{-\frac{\gamma_W}{\beta_0}}w(\mu)
\eeq
where $\beta_0=\frac{11N_c}{3}-\frac{2n_f}{3}$ and 
\beq
\gamma_W=\frac{N_c}{2}+n_f+\frac{\beta_0}{2} = \frac{7N_c}{3}+\frac{2n_f}{3}
\eeq
is the one-loop anomalous dimension of the Weinberg operator \cite{Morozov:1985ef}.\footnote{Recently, the anomalous dimension of ${\cal O}_W$ has been calculated to three-loops \cite{deVries:2019nsu}, but numerically this does not significantly affect the leading-order result. The corresponding three-loop results for the other entries of the anomalous dimension matrix are not available yet. } We immediately see that the product $wE$ that appears in (\ref{bigi}) is not RG-invariant, contrary to what one naively expects from the fact that the effective action $w\int d^4x {\cal O}_W(x)$ is RG-invariant. This is because we are dealing with nonforward matrix elements.  
We take $\mu$ to be the hadronic scale $\sim 1$ GeV, and $\mu_0 \gtrsim 100$ GeV is the high energy scale where the Weinberg operator is induced. 
Numerically, for $n_f=5$,  
\beq
\gamma_W\approx 10.3, \qquad \gamma_4\approx 6.89, \qquad   \frac{9N_c^2}{2(3N_c^2+4)} \approx 1.31.
\eeq
Since $\gamma_W>\gamma_4$, asymptotically $\mu_0\to \infty$ one has the relation
\beq
E(\mu_0)\approx 1.31 f_0(\mu_0), \label{asy}
\eeq
in the sense that the difference 
$|E(\mu_0) -  1.31 f_0(\mu_0)|$ is much smaller than $|f_0(\mu_0)|$ and $|E(\mu_0)|$. Eq.~(\ref{asy}) does not say anything definite about the relation between $E(\mu)$ and $f_0(\mu)$ at low energy, but it does suggest that $E(\mu)$ is in the ballpark of $1.3f_0(\mu)$ up to corrections of order unity.  Another insight into the relative coefficient may come from the exact operator identity \cite{Hatta:2020ltd}
\beq
{\cal O}_W = \partial_\mu(\bar{\psi}g\tilde{F}^{\mu\nu}\gamma_\nu \psi) -\frac{1}{2}\tilde{F}_{\mu\nu}D^2F^{\mu\nu} \equiv {\cal O}_4+{\cal O}_D,
\eeq
or in terms of matrix elements, 
\beq
E=\frac{f_0}{2} + \frac{1}{8im^2}\lim_{\Delta\to 0} \frac{
\langle P'|\tilde{F}_{\mu\nu}D^2F^{\mu\nu}|P\rangle}{\Delta \cdot S} \label{d}
\eeq
valid at any scale $\mu$. While we do not know the value of the matrix element $\langle \tilde{F}D^2F\rangle$, its scale dependence can be inferred from the evolution of ${\cal O}_D$  \cite{Hatta:2020ltd}
\beq
\frac{\partial {\cal O}_D}{\partial \ln \mu^2} =-\frac{\alpha_s}{4\pi}\left[\gamma_W {\cal O}_D  -\left(\frac{2N_c}{3}+\frac{8C_F}{3}\right){\cal O}_4\right]
\eeq
We see that, in addition to the strong suppression $\sim \gamma_W$ with increasing $\mu$, there is a compensating  contribution proportional to ${\cal O}_4\sim f_0$. This means that $E$ asymptotically  approaches $1.3f_0$  from below. We thus arrive at the following scenario.  At low energy,  $E(\mu)\sim 0.5f_0(\mu)$ simply from the first term in (\ref{d}). As $\mu$ increases, the ${\cal O}_4$ component of ${\cal O}_D$ grows due to mixing, and this eventually leads to $E\sim 1.3f_0$ at high energy. 
%(We remind the reader that the present normalization of $f_0$ differs from \cite{Hatta:2020ltd} by a factor of 2.) 
In other words, $E(\mu)$ sits somewhere between the two reference values 
\beq
0.5f_0(\mu) < E(\mu) < 1.3f_0(\mu). \label{model}
\eeq
We shall use this estimate to calculate nucleon EDMs in the next section. 

%\beq
%E(\mu)-\frac{9N_c^2}{2(3N_c^2+4)} f_0(\mu) =\left(\frac{\alpha_s(\mu_0)}{\alpha_s(\mu)}\right)^{-\frac{\gamma_W}{\beta_0}} \left(E(\mu_0)-\frac{9N_c^2}{2(3N_c^2+4)} f_0(\mu_0) \right)
%\eeq
% \beq
% f_0(\mu)= \left(\frac{\alpha_s(\mu_0)}{\alpha_s(\mu)}\right)^{-\frac{\gamma_4}{\beta_0}}  f_0(\mu_0), \qquad \gamma_4=  \frac{8C_F}{3}+\frac{2n_f}{3} \label{g4}
%\eeq

\section{Polarized DIS at twist-four and nucleon EDM}

Remarkably, the parameter $f_0$ is observable in experiments, as a part of higher twist corrections in polarized deep inelastic scattering (DIS).  
Up to twist-four accuracy, the first moment of the $g_1$ structure function for proton ($p$) and neutron ($n$) measurable in polarized DIS can be written as  \cite{Shuryak:1981pi,Balitsky:1989jb,Ji:1993sv,Kawamura:1996gg} 
\beq
\int_0^1 dx g_1^{p,n}(x) &=& \left(\pm \frac{g_A}{12} + \frac{a_8}{36}\right)\left(1-\frac{\alpha_s}{\pi}+{\cal O}(\alpha_s^2)\right) + \frac{\Delta \Sigma}{9}\left(1-\frac{33-8n_f}{33-2n_f}\frac{\alpha_s}{\pi} +{\cal O}(\alpha_s^2) \right) \nn 
&&+\frac{m^2}{9Q^2}(a_2^{p,n}+4d_2^{p,n}+4f^{p,n}_2),% +\frac{8m^2}{9Q^2}  \left(\pm \frac{f_3}{12} +\frac{f_8}{36} + \frac{f_0}{9}\right)
\label{twist}
\eeq
where we suppress the scale ($Q^2$) dependence of the low-energy constants for simplicity.\footnote{In DIS, it is common to choose $\mu^2=Q^2$, the virtuality of the photon emitted from the incoming lepton.} $a_3^p = -a_3^n = g_A$ is the isovector axial charge and we wrote $a_8^p=a_8^n\equiv a_8$ for the octet charge. $\Delta \Sigma$ is the quarks' helicity contribution to the nucleon spin. Usually, $\Delta \Sigma$ is the main object of interest as it is the crucial building block of the nucleon spin sum rule. Here, however,  we focus on the higher twist corrections in the second line of (\ref{twist}) which are suppressed by the characteristic factor $m^2/Q^2$.  The terms proportional to $a_2$ and $d_2$ are the so-called target mass corrections (TMCs) important in the low-$Q^2$ region. They originate from the finiteness of the nucleon mass and can be expressed by the third moments of the $g_1,g_2$ structure functions
\beq
a_2^{p,n}=2\int_0^1 dx x^2 g^{p,n}_1(x), \qquad d_2^{p,n}= \int_0^1 dx x^2 (2g^{p,n}_1(x)+3g^{p,n}_2(x)). \label{three}
\eeq
 Eq.~(\ref{three}) shows that the TMCs are related to twist-two and twist-three matrix elements, although in practice they  appear as twist-four effects $\sim 1/Q^2$ in the cross section.  
 The `genuine' twist-four contribution can be written as 
 \beq
 f_2^{p,n}= 2  \left(\pm \frac{f_3}{12} +\frac{f_8}{36} + \frac{f_0}{9}\right) \label{ff}
 \eeq
 where  $f_{0,3,8}$ are defined through the following matrix elements
\beq
\langle PS|\bar{\psi} g\tilde{F}^{\mu\nu}\gamma_\nu  T^a\psi|PS\rangle  =-2f_a^{p,n} m^2 S^\mu \label{min}
\eeq
with $T^{a=0,3,8}$ being $3\times 3$ flavor matrices 
\beq
T^0=\mathbb{1}, \qquad T^3={\rm diag}(1,-1,0), \qquad T^8={\rm diag}(1,1,-2).
\eeq
%\beq
%f_0=f_{2u}+f_{2d}+f_{2s},  \qquad f_3=f_{2u}-f_{2d}, \qquad f_{8}=f_{2u}+f_{2d}-2f_{2s}
%\eeq
The minus sign in (\ref{min}), which is absent in most literature, is due to our sign convention of the coupling $g$, namely  $D^\mu=\partial^\mu+igA^\mu$. 
In (\ref{ff}), we wrote $f_{0,8}^p=f_{0,8}^n\equiv f_{0,8}$ and $f_3^p=-f_3^n \equiv f_3$. Note that $f_0$ is the same as in the previous definition (\ref{same}). 

There have already been several attempts to extract the twist-four contributions from the world data of polarized DIS \cite{Ji:1997gs,Leader:2002ni,Meziani:2004ne,Leader:2006xc}.
The original result in \cite{Ji:1997gs} is
\beq
f_2^p=0.10\pm 0.05, \qquad f_2^n =0.07\pm 0.08 \qquad (Q^2=1\, {\rm GeV}^2)
\eeq
(The sign convention of $f_2$ in  \cite{Ji:1997gs} is opposite to ours, so we flipped the sign.) However, the precision of the early data used in this reference was not very good. 
The more recent  results for the neutron based on experiments at the Jefferson Lab are\footnote{Ref.~\cite{Leader:2006xc} also studied the proton case, but the authors only extracted the sum $d_2^{p,n}+f_2^{p,n}$.} 
\beq
&&{\rm Ref.[20]} \qquad f_2^n = 0.034\pm 0.043\qquad  (Q^2=1\, {\rm GeV}^2) \nn
&&{\rm Ref.[22]} \qquad f_2^n = 0.053\pm 0.026 \qquad (Q^2=3.21\, {\rm GeV}^2) \label{j}
\eeq
Once $f_2^{p,n}$ are known, one can extract the linear combination 
\beq
\frac{f_8(Q^2)}{36}  + \frac{f_0(Q^2)}{9} =\frac{f_8(Q_0^2)}{36} \left(\frac{\alpha_s(Q_0^2)}{\alpha_s (Q^2)}\right)^{-\gamma_4^{NS}/\beta_0}  + \frac{f_0(Q_0^2)}{9}\left(\frac{\alpha_s(Q_0^2)}{\alpha_s (Q^2)}\right)^{-\gamma_4/\beta_0} \label{lin}
\eeq
 where  $\gamma_4^{NS}=\frac{8C_F}{3}$ and $\gamma_4$ is the same as in (\ref{g4}). This in principle allows one to separately determine  $f_0$ and $f_8$. However, in order to do so,  very accurate data over a very wide range in $Q^2$ are needed.  Since such data are not available at the moment,  we resort to model calculations of $f_a$ from QCD sum rules \cite{Balitsky:1989jb,Stein:1995si} and instantons
 \cite{Balla:1997hf,Lee:2001ug}. Here we quote the result of \cite{Lee:2001ug} since the predicted value  $f_2^n$ agrees well  with (\ref{j}):
\beq
&&f_0 = 0.01, \qquad f_3=-0.25, \qquad f_8=-0.11\nn
&& f_2^p=-0.046, \qquad f_2^n = 0.038 \label{ch}
\eeq
 at the defining scale of the model  $Q^2=0.4$ GeV$^2$. The isoscalar form factor $f_0$ is small partly because it is subleading in the large-$N_c$ counting $f_0\sim 1/N_c$. The theoretical uncertainties of the result (\ref{ch}) come from the validity of the dilute instanton approximation as well as that of the $1/N_c$ expansion which are difficult to quantify especially in the flavor singlet sector. A conservative estimate for $f_0$ is a  factor of 2 uncertainties in both directions \cite{weiss}.      
 %The large value of $f_3$ may seem worrisome, but it cancels in the isosinglet combination $f_2^p+f_2^n$. 
 %Given  the difference in the prefactors, $1/36$ and $1/9$, we see that in this model the contribution to (\ref{lin})  from $f_8$  is larger than that from $f_0$ by a factor of $2\sim3$. 

We are now ready to present numerical results. Varying $E$ in the range (\ref{model}) and  using $f_0=0.01$ and the known constants\footnote{Note that we use the total magnetic moment of the proton as explained earlier. }
\beq
\mu_p=2.79\mu_N, \qquad 
\mu_n=-1.91\mu_N, \qquad 
\mu_N=0.105 \, e\, {\rm fm}, \qquad 
m^2 \mu_N \approx 470\,e\,{\rm MeV},
\eeq
 we find\footnote{We do not know the sign of $w$. Here we assume it is positive, but if it turns out to be negative, the inequality symbols should be modified accordingly. } 
\beq
26w\, e\,{\rm MeV}< d_p < 69w\, e\,{\rm MeV}, \qquad -47w\, e \,{\rm MeV} <d_n< -18w\, e \,{\rm MeV}, \label{main}
\eeq
Note that this does not include the  uncertainty of $f_0$ mentioned above. 
Let us  compare (\ref{main}) with the previous results based on QCD sum rules \cite{Demir:2002gg,Haisch:2019bml}. The results in these two references are consistent with each other, so we only quote the result of \cite{Haisch:2019bml}. Our normalization of ${\cal O}_W$ differs from \cite{Haisch:2019bml}  by a factor of $g$, and our sign convention of  $g$ is opposite. Since Ref.~\cite{Haisch:2019bml} assumed $g_{[9]}=2.13$ is positive, we have to divide the numbers in (\ref{main}) by  $g=-g_{[9]}=- 2.13$ to get  ($w'\equiv gw$)
%\beq
%-32w' \, e\,{\rm MeV}< d_p <-12 w' \, e\,{\rm MeV}   \qquad 8.4 w' \,e\,{\rm MeV}< d_n <22w'\, e \,{\rm MeV}. \label{mag}
%\eeq
\beq
-12 w' \, e\,{\rm MeV} < d_p < -32w' \, e\,{\rm MeV}   \qquad 22w'\, e \,{\rm MeV}< d_n<8.4 w' \,e\,{\rm MeV}. \label{mag}
\eeq
This should be compared to   \cite{Haisch:2019bml}
\beq
d_p = -109(1\pm 0.5)  \,e\,{\rm MeV}, \qquad d_n=74(1\pm 0.5)  \,e\,{\rm MeV}. \label{ha}
\eeq
($w'$ has been set to unity in \cite{Haisch:2019bml}.)   
The signs agree, but if we compare the central values, our results (\ref{mag}) are smaller than (\ref{ha}) by a factor of about 3 even in the maximal scenario $E=1.3f_0$ and this factor becomes $8\sim 9$ in the minimal case $E=0.5f_0$. The suppression is mainly attributed  to the small twist-four matrix element $f_0\sim 0.01$ \cite{Lee:2001ug}. However, it should be kept in mind  that  theoretical uncertainties in these (instanton/sum rule) calculations are large. In fact, if we pick up the minus sign in (\ref{ha}) and consider the  upward uncertainty of $f_0$, the two results are not entirely inconsistent. 

\section{Conclusions}

In this paper, we have presented a novel prediction for the nucleon EDMs originating from the Weinberg operator. This has been  made possible owing to the recent  observation \cite{Hatta:2020ltd} that the matrix element of the Weinberg operator is related to the parameter $f_0$ that can be  extracted from polarized DIS experiments. The main result is shown in Eq.~(\ref{main}). Admittedly, our estimate is rather crude and subject to the  uncertainties in the calculation of $f_0$ (\ref{ch}), not to mention the uncertainties associated with (\ref{model}). While these theoretical issues could be improved in future, ideally  $f_0$ should be extracted from polarized DIS experiments, and  this is our main message after all. In order to reliably extract $f_0$,  we need very precise data over a wide range in $Q^2$ (see  (\ref{lin})), while the existing data are rather limited in $Q^2$. The future polarized DIS experiments at the EIC in the US.  and the EIcC (EIC in China)  may help  in this regard. Hopefully our work triggers discussions on the  feasibility of such studies in these experiments. 

Concerning the result, the obtained values $d_{p,n}$ tend to be  smaller (by a factor of $3\sim 9$) than the previous results based on QCD sum rules  \cite{Demir:2002gg,Haisch:2019bml}, although they are not entirely inconsistent after including large theoretical uncertainties. It is difficult to make a detailed step-by-step comparison, since the method used in \cite{Demir:2002gg,Haisch:2019bml} is very   different from ours. To better understand this, it may be helpful to revisit the sum rule calculations of $f_0$  \cite{Balitsky:1989jb,Stein:1995si}.  

As a matter of fact, numerically our result is comparable to the contribution from the one-nucleon irreducible diagram Fig.~\ref{fig1}(c) recently calculated in \cite{Yamanaka:2020kjo}.\footnote{To compare with the result of \cite{Yamanaka:2020kjo}, the numbers in (\ref{mag}) have to be divided by 3.}  This calls in question the original argument that the  reducible diagrams dominate over the irreducible ones. However, as we already mentioned, all these phenomenological estimates contain large systematic errors. Further efforts in both theory and experiment are certainly needed to draw definitive conclusions.

\section*{Acknowledgments}
I am grateful to Nodoka Yamanaka, Elliot Leader and Christian Weiss for correspondence. I thank the Yukawa Institute for Theoretical Physics for hospitality. 
This work is supported by the U.S. Department of Energy, Office of
Science, Office of Nuclear Physics, under contract No. DE- SC0012704,
and in part by Laboratory Directed Research and Development (LDRD)
funds from Brookhaven Science Associates.

%\bibliography{references}

\begin{thebibliography}{24}
\expandafter\ifx\csname natexlab\endcsname\relax\def\natexlab#1{#1}\fi
\expandafter\ifx\csname bibnamefont\endcsname\relax
  \def\bibnamefont#1{#1}\fi
\expandafter\ifx\csname bibfnamefont\endcsname\relax
  \def\bibfnamefont#1{#1}\fi
\expandafter\ifx\csname citenamefont\endcsname\relax
  \def\citenamefont#1{#1}\fi
\expandafter\ifx\csname url\endcsname\relax
  \def\url#1{\texttt{#1}}\fi
\expandafter\ifx\csname urlprefix\endcsname\relax\def\urlprefix{URL }\fi
\providecommand{\bibinfo}[2]{#2}
\providecommand{\eprint}[2][]{\url{#2}}

\bibitem[{\citenamefont{Hatta}(2020)}]{Hatta:2020ltd}
\bibinfo{author}{\bibfnamefont{Y.}~\bibnamefont{Hatta}},
  \bibinfo{journal}{Phys. Rev. D} \textbf{\bibinfo{volume}{102}},
  \bibinfo{pages}{094004} (\bibinfo{year}{2020}), \eprint{2009.03657}.

\bibitem[{\citenamefont{Chupp et~al.}(2019)\citenamefont{Chupp, Fierlinger,
  Ramsey-Musolf, and Singh}}]{Chupp:2017rkp}
\bibinfo{author}{\bibfnamefont{T.}~\bibnamefont{Chupp}},
  \bibinfo{author}{\bibfnamefont{P.}~\bibnamefont{Fierlinger}},
  \bibinfo{author}{\bibfnamefont{M.}~\bibnamefont{Ramsey-Musolf}},
  \bibnamefont{and} \bibinfo{author}{\bibfnamefont{J.}~\bibnamefont{Singh}},
  \bibinfo{journal}{Rev. Mod. Phys.} \textbf{\bibinfo{volume}{91}},
  \bibinfo{pages}{015001} (\bibinfo{year}{2019}), \eprint{1710.02504}.

\bibitem[{\citenamefont{Yamanaka et~al.}(2017)\citenamefont{Yamanaka, Sahoo,
  Yoshinaga, Sato, Asahi, and Das}}]{Yamanaka:2017mef}
\bibinfo{author}{\bibfnamefont{N.}~\bibnamefont{Yamanaka}},
  \bibinfo{author}{\bibfnamefont{B.}~\bibnamefont{Sahoo}},
  \bibinfo{author}{\bibfnamefont{N.}~\bibnamefont{Yoshinaga}},
  \bibinfo{author}{\bibfnamefont{T.}~\bibnamefont{Sato}},
  \bibinfo{author}{\bibfnamefont{K.}~\bibnamefont{Asahi}}, \bibnamefont{and}
  \bibinfo{author}{\bibfnamefont{B.}~\bibnamefont{Das}}, \bibinfo{journal}{Eur.
  Phys. J. A} \textbf{\bibinfo{volume}{53}}, \bibinfo{pages}{54}
  (\bibinfo{year}{2017}), \eprint{1703.01570}.

\bibitem[{\citenamefont{Weinberg}(1989)}]{Weinberg:1989dx}
\bibinfo{author}{\bibfnamefont{S.}~\bibnamefont{Weinberg}},
  \bibinfo{journal}{Phys. Rev. Lett.} \textbf{\bibinfo{volume}{63}},
  \bibinfo{pages}{2333} (\bibinfo{year}{1989}).

\bibitem[{\citenamefont{Cirigliano et~al.}(2020)\citenamefont{Cirigliano,
  Mereghetti, and Stoffer}}]{Cirigliano:2020msr}
\bibinfo{author}{\bibfnamefont{V.}~\bibnamefont{Cirigliano}},
  \bibinfo{author}{\bibfnamefont{E.}~\bibnamefont{Mereghetti}},
  \bibnamefont{and} \bibinfo{author}{\bibfnamefont{P.}~\bibnamefont{Stoffer}},
  \bibinfo{journal}{JHEP} \textbf{\bibinfo{volume}{09}}, \bibinfo{pages}{094}
  (\bibinfo{year}{2020}), \eprint{2004.03576}.

\bibitem[{\citenamefont{Rizik et~al.}(2020)\citenamefont{Rizik, Monahan, and
  Shindler}}]{Rizik:2020naq}
\bibinfo{author}{\bibfnamefont{M.~D.} \bibnamefont{Rizik}},
  \bibinfo{author}{\bibfnamefont{C.~J.} \bibnamefont{Monahan}},
  \bibnamefont{and} \bibinfo{author}{\bibfnamefont{A.}~\bibnamefont{Shindler}}
  (\bibinfo{collaboration}{SymLat}), \bibinfo{journal}{Phys. Rev. D}
  \textbf{\bibinfo{volume}{102}}, \bibinfo{pages}{034509}
  (\bibinfo{year}{2020}), \eprint{2005.04199}.

\bibitem[{\citenamefont{Bigi and Uraltsev}(1991)}]{Bigi:1990kz}
\bibinfo{author}{\bibfnamefont{I.~I.} \bibnamefont{Bigi}} \bibnamefont{and}
  \bibinfo{author}{\bibfnamefont{N.}~\bibnamefont{Uraltsev}},
  \bibinfo{journal}{Nucl. Phys. B} \textbf{\bibinfo{volume}{353}},
  \bibinfo{pages}{321} (\bibinfo{year}{1991}).

\bibitem[{\citenamefont{Demir et~al.}(2003)\citenamefont{Demir, Pospelov, and
  Ritz}}]{Demir:2002gg}
\bibinfo{author}{\bibfnamefont{D.~A.} \bibnamefont{Demir}},
  \bibinfo{author}{\bibfnamefont{M.}~\bibnamefont{Pospelov}}, \bibnamefont{and}
  \bibinfo{author}{\bibfnamefont{A.}~\bibnamefont{Ritz}},
  \bibinfo{journal}{Phys. Rev. D} \textbf{\bibinfo{volume}{67}},
  \bibinfo{pages}{015007} (\bibinfo{year}{2003}), \eprint{hep-ph/0208257}.

\bibitem[{\citenamefont{Haisch and Hala}(2019)}]{Haisch:2019bml}
\bibinfo{author}{\bibfnamefont{U.}~\bibnamefont{Haisch}} \bibnamefont{and}
  \bibinfo{author}{\bibfnamefont{A.}~\bibnamefont{Hala}},
  \bibinfo{journal}{JHEP} \textbf{\bibinfo{volume}{11}}, \bibinfo{pages}{154}
  (\bibinfo{year}{2019}), \eprint{1909.08955}.

\bibitem[{\citenamefont{Yamanaka and Hiyama}(2020)}]{Yamanaka:2020kjo}
\bibinfo{author}{\bibfnamefont{N.}~\bibnamefont{Yamanaka}} \bibnamefont{and}
  \bibinfo{author}{\bibfnamefont{E.}~\bibnamefont{Hiyama}}
  (\bibinfo{year}{2020}), \eprint{2011.02531}.

\bibitem[{\citenamefont{Prokudin et~al.}(2020)\citenamefont{Prokudin, Hatta,
  Kovchegov, and Marquet}}]{Proceedings:2020eah}
\bibinfo{editor}{\bibfnamefont{A.}~\bibnamefont{Prokudin}},
  \bibinfo{editor}{\bibfnamefont{Y.}~\bibnamefont{Hatta}},
  \bibinfo{editor}{\bibfnamefont{Y.}~\bibnamefont{Kovchegov}},
  \bibnamefont{and} \bibinfo{editor}{\bibfnamefont{C.}~\bibnamefont{Marquet}},
  eds., \emph{\bibinfo{title}{{Proceedings, Probing Nucleons and Nuclei in High
  Energy Collisions: Dedicated to the Physics of the Electron Ion Collider}:
  {Seattle (WA), United States, October 1 - November 16, 2018}}}
  (\bibinfo{publisher}{WSP}, \bibinfo{year}{2020}), \eprint{2002.12333}.

\bibitem[{\citenamefont{Morozov}(1984)}]{Morozov:1985ef}
\bibinfo{author}{\bibfnamefont{A.}~\bibnamefont{Morozov}},
  \bibinfo{journal}{Sov. J. Nucl. Phys.} \textbf{\bibinfo{volume}{40}},
  \bibinfo{pages}{505} (\bibinfo{year}{1984}).

%\cite{deVries:2019nsu}
\bibitem{deVries:2019nsu}
J.~de Vries, G.~Falcioni, F.~Herzog and B.~Ruijl,
%``Two- and three-loop anomalous dimensions of Weinberg\textquoteright{}s dimension-six CP-odd gluonic operator,''
Phys. Rev. D \textbf{102}, 016010 (2020),
%doi:10.1103/PhysRevD.102.016010
1907.04923.
%10 citations counted in INSPIRE as of 06 Dec 2020



\bibitem[{\citenamefont{Shuryak and Vainshtein}(1982)}]{Shuryak:1981pi}
\bibinfo{author}{\bibfnamefont{E.~V.} \bibnamefont{Shuryak}} \bibnamefont{and}
  \bibinfo{author}{\bibfnamefont{A.}~\bibnamefont{Vainshtein}},
  \bibinfo{journal}{Nucl. Phys. B} \textbf{\bibinfo{volume}{201}},
  \bibinfo{pages}{141} (\bibinfo{year}{1982}).

\bibitem[{\citenamefont{Balitsky et~al.}(1990)\citenamefont{Balitsky, Braun,
  and Kolesnichenko}}]{Balitsky:1989jb}
\bibinfo{author}{\bibfnamefont{I.}~\bibnamefont{Balitsky}},
  \bibinfo{author}{\bibfnamefont{V.~M.} \bibnamefont{Braun}}, \bibnamefont{and}
  \bibinfo{author}{\bibfnamefont{A.}~\bibnamefont{Kolesnichenko}},
  \bibinfo{journal}{Phys. Lett. B} \textbf{\bibinfo{volume}{242}},
  \bibinfo{pages}{245} (\bibinfo{year}{1990}), \bibinfo{note}{[Erratum:
  Phys.Lett.B 318, 648 (1993)]}, \eprint{hep-ph/9310316}.

\bibitem[{\citenamefont{Ji and Unrau}(1994)}]{Ji:1993sv}
\bibinfo{author}{\bibfnamefont{X.-D.} \bibnamefont{Ji}} \bibnamefont{and}
  \bibinfo{author}{\bibfnamefont{P.}~\bibnamefont{Unrau}},
  \bibinfo{journal}{Phys. Lett. B} \textbf{\bibinfo{volume}{333}},
  \bibinfo{pages}{228} (\bibinfo{year}{1994}), \eprint{hep-ph/9308263}.

\bibitem[{\citenamefont{Kawamura et~al.}(1997)\citenamefont{Kawamura, Uematsu,
  Kodaira, and Yasui}}]{Kawamura:1996gg}
\bibinfo{author}{\bibfnamefont{H.}~\bibnamefont{Kawamura}},
  \bibinfo{author}{\bibfnamefont{T.}~\bibnamefont{Uematsu}},
  \bibinfo{author}{\bibfnamefont{J.}~\bibnamefont{Kodaira}}, \bibnamefont{and}
  \bibinfo{author}{\bibfnamefont{Y.}~\bibnamefont{Yasui}},
  \bibinfo{journal}{Mod. Phys. Lett. A} \textbf{\bibinfo{volume}{12}},
  \bibinfo{pages}{135} (\bibinfo{year}{1997}), \eprint{hep-ph/9603338}.

\bibitem[{\citenamefont{Ji and Melnitchouk}(1997)}]{Ji:1997gs}
\bibinfo{author}{\bibfnamefont{X.-D.} \bibnamefont{Ji}} \bibnamefont{and}
  \bibinfo{author}{\bibfnamefont{W.}~\bibnamefont{Melnitchouk}},
  \bibinfo{journal}{Phys. Rev. D} \textbf{\bibinfo{volume}{56}},
  \bibinfo{pages}{1} (\bibinfo{year}{1997}), \eprint{hep-ph/9703363}.

\bibitem[{\citenamefont{Leader et~al.}(2003)\citenamefont{Leader, Sidorov, and
  Stamenov}}]{Leader:2002ni}
\bibinfo{author}{\bibfnamefont{E.}~\bibnamefont{Leader}},
  \bibinfo{author}{\bibfnamefont{A.~V.} \bibnamefont{Sidorov}},
  \bibnamefont{and} \bibinfo{author}{\bibfnamefont{D.~B.}
  \bibnamefont{Stamenov}}, \bibinfo{journal}{Phys. Rev. D}
  \textbf{\bibinfo{volume}{67}}, \bibinfo{pages}{074017}
  (\bibinfo{year}{2003}), \eprint{hep-ph/0212085}.

\bibitem[{\citenamefont{Meziani et~al.}(2005)}]{Meziani:2004ne}
\bibinfo{author}{\bibfnamefont{Z.}~\bibnamefont{Meziani}} \bibnamefont{et~al.},
  \bibinfo{journal}{Phys. Lett. B} \textbf{\bibinfo{volume}{613}},
  \bibinfo{pages}{148} (\bibinfo{year}{2005}), \eprint{hep-ph/0404066}.

\bibitem[{\citenamefont{Leader et~al.}(2007)\citenamefont{Leader, Sidorov, and
  Stamenov}}]{Leader:2006xc}
\bibinfo{author}{\bibfnamefont{E.}~\bibnamefont{Leader}},
  \bibinfo{author}{\bibfnamefont{A.~V.} \bibnamefont{Sidorov}},
  \bibnamefont{and} \bibinfo{author}{\bibfnamefont{D.~B.}
  \bibnamefont{Stamenov}}, \bibinfo{journal}{Phys. Rev. D}
  \textbf{\bibinfo{volume}{75}}, \bibinfo{pages}{074027}
  (\bibinfo{year}{2007}), \eprint{hep-ph/0612360}.

\bibitem[{\citenamefont{Flay et~al.}(2016)}]{Flay:2016wie}
\bibinfo{author}{\bibfnamefont{D.}~\bibnamefont{Flay}} \bibnamefont{et~al.}
  (\bibinfo{collaboration}{Jefferson Lab Hall A}), \bibinfo{journal}{Phys. Rev.
  D} \textbf{\bibinfo{volume}{94}}, \bibinfo{pages}{052003}
  (\bibinfo{year}{2016}), \eprint{1603.03612}.

\bibitem[{\citenamefont{Stein et~al.}(1995)\citenamefont{Stein, Gornicki,
  Mankiewicz, and Schafer}}]{Stein:1995si}
\bibinfo{author}{\bibfnamefont{E.}~\bibnamefont{Stein}},
  \bibinfo{author}{\bibfnamefont{P.}~\bibnamefont{Gornicki}},
  \bibinfo{author}{\bibfnamefont{L.}~\bibnamefont{Mankiewicz}},
  \bibnamefont{and} \bibinfo{author}{\bibfnamefont{A.}~\bibnamefont{Schafer}},
  \bibinfo{journal}{Phys. Lett. B} \textbf{\bibinfo{volume}{353}},
  \bibinfo{pages}{107} (\bibinfo{year}{1995}), \eprint{hep-ph/9502323}.

\bibitem[{\citenamefont{Balla et~al.}(1998)\citenamefont{Balla, Polyakov, and
  Weiss}}]{Balla:1997hf}
\bibinfo{author}{\bibfnamefont{J.}~\bibnamefont{Balla}},
  \bibinfo{author}{\bibfnamefont{M.~V.} \bibnamefont{Polyakov}},
  \bibnamefont{and} \bibinfo{author}{\bibfnamefont{C.}~\bibnamefont{Weiss}},
  \bibinfo{journal}{Nucl. Phys. B} \textbf{\bibinfo{volume}{510}},
  \bibinfo{pages}{327} (\bibinfo{year}{1998}), \eprint{hep-ph/9707515}.

\bibitem[{\citenamefont{Lee et~al.}(2002)\citenamefont{Lee, Goeke, and
  Weiss}}]{Lee:2001ug}
\bibinfo{author}{\bibfnamefont{N.-Y.} \bibnamefont{Lee}},
  \bibinfo{author}{\bibfnamefont{K.}~\bibnamefont{Goeke}}, \bibnamefont{and}
  \bibinfo{author}{\bibfnamefont{C.}~\bibnamefont{Weiss}},
  \bibinfo{journal}{Phys. Rev. D} \textbf{\bibinfo{volume}{65}},
  \bibinfo{pages}{054008} (\bibinfo{year}{2002}), \eprint{hep-ph/0105173}.

\bibitem{weiss}
C.~Weiss, private communications.

\end{thebibliography}

\end{document}